# Preparation of Highly Crystalline TiO$_2$ Nanostructures by Acid-assisted Hydrothermal Treatment of Hexagonal-structured Nanocrystalline Titania/Cetyltrimethyammonium Bromide Nanoskeleton

Shuxi Dai 1, 2, Yanqiang Wu 1, Toshio Sakai 3, Zuliang Du 1 ✉, Hideki Sakai 2 ✉ and Masahiko Abe 2

(1) Key Laboratory for Special Functional Materials of Ministry of Education, Henan University, Kaifeng, 475004, People's Republic of China
(2) Department of Pure and Applied Chemistry, Faculty of Science and Technology, Tokyo University of Science, Chiba, Noda 278-8510, Japan
(3) Internaltional Young Researchers Empowerment Center, Shinshu University, Wakasato, Nagano 380-8553, Japan

**Abstract** Highly crystalline TiO$_2$ nanostructures were prepared through a facile inorganic acid-assisted hydrothermal treatment of hexagonal-structured assemblies of nanocrystalline titania templated by cetyltrimethylammonium bromide (Hex-ncTiO$_2$/CTAB Nanoskeleton) as starting materials. All samples were characterized by X-ray diffraction (XRD) and transmission electron microscopy (TEM). The influence of hydrochloric acid concentration on the morphology, crystalline and the formation of the nanostructures were investigated. We found that the morphology and crystalline phase strongly depended on the hydrochloric acid concentrations. More importantly, crystalline phase was closely related to the morphology of TiO$_2$ nanostructure. Nanoparticles were polycrystalline anatase phase, and aligned nanorods were single crystalline rutile phase. Possible formation mechanisms of TiO$_2$ nanostructures with various crystalline phases and morphologies were proposed.

S. Dai · Y. Wu · Z. Du (✉)
Key Laboratory for Special Functional Materials of Ministry of Education, Henan University, Kaifeng 475004, People's Republic of China
e-mail: zld@henu.edu.cn

H. Sakai (✉) · M. Abe
Department of Pure and Applied Chemistry, Faculty of Science and Technology, Tokyo University of Science, Chiba, Noda 278-8510, Japan
e-mail: hisakai@rs.noda.tus.ac.jp

T. Sakai
Internaltional Young Researchers Empowerment Center, Shinshu University, Wakasato, Nagano 380-8553, Japan

## Introduction

Titanium oxide (TiO$_2$) is an important semiconductor material for use in a wide range of applications, including photocatalysis, environmental pollution control and solar energy conversion [1–4]. It is well known that titanium dioxide exists in three crystalline polymorphs, namely rutile (tetragonal), anatase (tetragonal), and brookite (orthorhombic). Rutile is the most stable phase, whereas anatase and brookite are metastable phase and transform to rutile upon heating [5, 7]. The rutile phase has been widely used for pigment materials because of its chemical stability. However, the anatase phase has been widely used in photodegradation due to its high photoactivity [1, 4, 6–8]. The majority of the applications of TiO$_2$ are strongly influenced by the crystalline phase [9]. In order to obtain highly crystalline TiO$_2$ at low temperature, a long time is necessary to get pure anatase or anatase–rutile phase mixture, and days or even longer time for the formation of rutile phase in traditional sol–gel process [10–12]. Anatase or the mixture of anatase and rutile can be produced by calcination within several hours with use of amorphous TiO$_2$ as starting materials. However, obtaining pure anatase requires calcinations at 500°C, and pure rutile needs higher temperature [12], which often resulted in the collapse of the unique nanostructures, such as nanotube, nanorods, and formation of a conglomeration influencing the applications of TiO$_2$. How to synthesize highly crystalline TiO$_2$ at a relative lower temperature is still a difficult and hot topic in recent years [13, 14].

In comparison with sol–gel method [15, 16], hydrothermal synthesis is an easy route to prepare a well-crystalline oxide under the moderate reaction condition, i.e. low temperature and short reaction time [17]. Hydrothermal media provides an effective reaction environment for the synthesis of nanocrystalline $TiO_2$ with high purity, good dispersion and well-controlled crystalline. The reactivity of a precursor system can be judged only by optimizing the processing variables such as starting materials, pH, and temperature [17]. To take advantage of the opportunities offered by hydrothermal synthesis, it is important to select a proper precursor system that is both reactive and cost effective. Recently, nanocrystalline $TiO_2$ particles with different structures and morphologies have been synthesized in hydrothermal media using different starting materials such as $TiCl_4$ [13, 18], $TiCl_3$ [7, 19, 20], amorphous $TiO_2$ [28], P25 [21], and titanate hydrates [1]. However, the preparation process of such starting materials is relatively complicated and the precursors are usually expensive and unstable.

In our previous work [16], we chose titanium oxysulfate sulfuric acid hydrate ($TiOSO_4 \cdot xH_2SO_4 \cdot xH_2O$) as a titania precursor and cetyltrimethylammonium bromide (CTAB) as a structure-directing agent for the preparation of titania. Both $TiOSO_4$ and CTAB are cheap and common materials for industries. After simply mixed together at a lower range of temperatures (30–60°C), hexagonal-structured assemblies of nanocrystalline titania were formed through hydrolysis of $TiOSO_4$ promoted by CTAB spherical micelles and condensation process (named as Hex-nc$TiO_2$/CTAB nanoskeleton) [22, 23]. This system had some unique features and advantages, for example, a facile preparation, crystallization of titania in aqueous solution in mild conditions and formation of hexagonal-structured anatase titania framework. Then, we were successful to prepare mesoporous titania particles with honeycomb structure and anatase crystalline framework after the calcinations of the Hex-nc$TiO_2$/CTAB nanoskeleton at 723 K for 2 h [16, 23].

In this paper, we examined the preparation of highly crystalline titanium dioxide nanostructures from the acid-assisted hydrothermal treatment of the Hex-nc$TiO_2$/CTAB nanoskeleton as a starting material. Nanostructured $TiO_2$ with different crystalline phases, crystallinity, and morphologies were obtained. In addition, the effect of hydrochloric acid concentration on the evolution of crystalline structure and morphologies of nanostructural $TiO_2$ products were investigated.

**Experimental Section**

Cetyltrimethylammonium bromide (CTAB) (Sigma, USA) was used as template material. Titanium oxysulfate sulfuric acid complex hydrate ($TiOSO_4 \cdot xH_2SO_4 \cdot xH_2O$) (Aldrich USA) was used as titania precursor. Hydrochloric acid (HCl) (Luoyang Chemical Reagents Factory, China) aqueous solutions were used as solvents in the hydrothermal procedure.

CTAB/$TiO_2$ hexagonal structures were prepared in the following procedures [16, 22, 23]. A concentration of 2.4 g $TiOSO_4$ was mixed with 25 mL $H_2O$ under constant magnetic stirring until the mixed solution turned into colorless solution at 50°C, and then 25 mL CTAB (60 mM) was added into the colorless solution and hold statically for 12 h at 50°C. The product obtained was filtered, washed with distilled water for several times, and dried at 120°C overnight.

Hydrochloric acid aqueous solutions with different concentrations were initially prepared from concentrated HCl with distilled $H_2O$, including 0.1–8 M. Subsequently, 0.5 g Hex-nc$TiO_2$/CTAB nanoskeleton was dispersed in 30 mL of the HCl aqueous solutions with stirring for 0.5 h, and then transferred into 50-mL container of a Teflon-lined stainless steel autoclave. The autoclave was heated and maintained at 150°C for 24 h and then cooled to room temperature. The precipitate was collected, centrifuged, washed with distilled water for several times, and then dried in a vacuum oven overnight at 60°C.

XRD patterns of the samples were collected with a Philips X' Pert Pro MPD X-ray diffraction system (XRD, Cu-K$\alpha$ radiation, $\lambda = 0.154056$ nm). All the samples were measured in the continuous scan mode in the $2\theta$ range of 10–90°, using a scan rate of 0.02 deg/s. The crystallite size was calculated using the Scherrer equation [25]. The morphology and structure of the products were observed with transmission electron microscopy (TEM) by JEM-2010 (JEOL Corporation, Japan), operating at 200 kV. The optical absorption spectra were obtained with Lambda 35 UV–vis spectrometer (Perkin-Elmer Inc., USA). $BaSO_4$ was used as a reflectance standard in the UV–visible diffuse reflectance experiment.

**Results and Discussion**

Figure 1 shows the typical TEM images of the Hex-nc$TiO_2$/CTAB nanoskeleton that we obtained. The Hex-nc$TiO_2$/CTAB nanoskeleton possessed long-range order with a hexagonal honeycomb structure and the pore size was 4.5 nm approximately, the thickness of the inorganic framework composed of titanium dioxide particles was about 1 nm. The low-angle XRD pattern in Fig. 2a shows that three diffraction peaks ($2\theta = 2.2°, 3.8°, 4.2°$) can be assigned to the long-range hexagonal structure of CTAB/$TiO_2$ mixture (d100:d110:d200=1:1/$\sqrt{3}$:1/2). The distance between pores was 4.4 nm calculated by Bragg' equation ($2d\sin\theta = \lambda$, $\lambda = 1.54056$ Å), which coincided with TEM data. Figure 2b shows the wide-angle XRD diffraction pattern of Hex-ncTiO$_2$/CTAB nanoskeleton. The pattern exhibited two primary diffraction peaks ($2\theta$ = 25.4, 48.0°) that can be assigned to the anatase phase structure. The weak and broadened diffraction peaks indicated that the Hex-ncTiO$_2$/CTAB nanoskeletons were poorly crystallized anatase structure and partly amorphous.

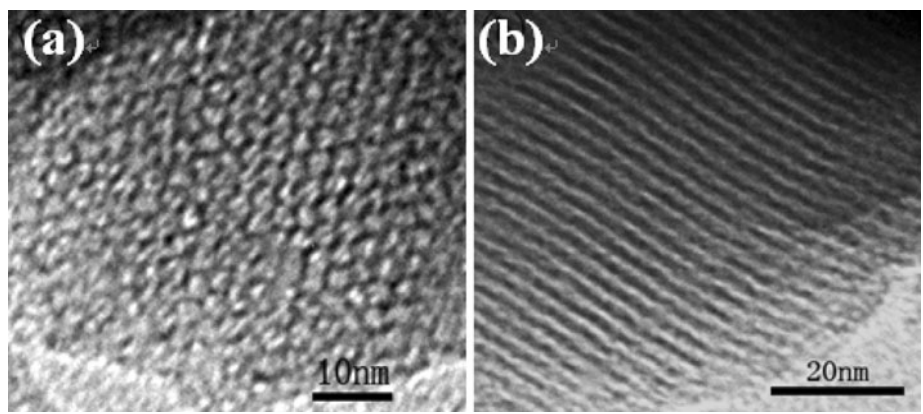

**Fig. 1** TEM images of Hex-ncTiO$_2$/CTAB Nanoskeleton. **a** Top view **b** side view

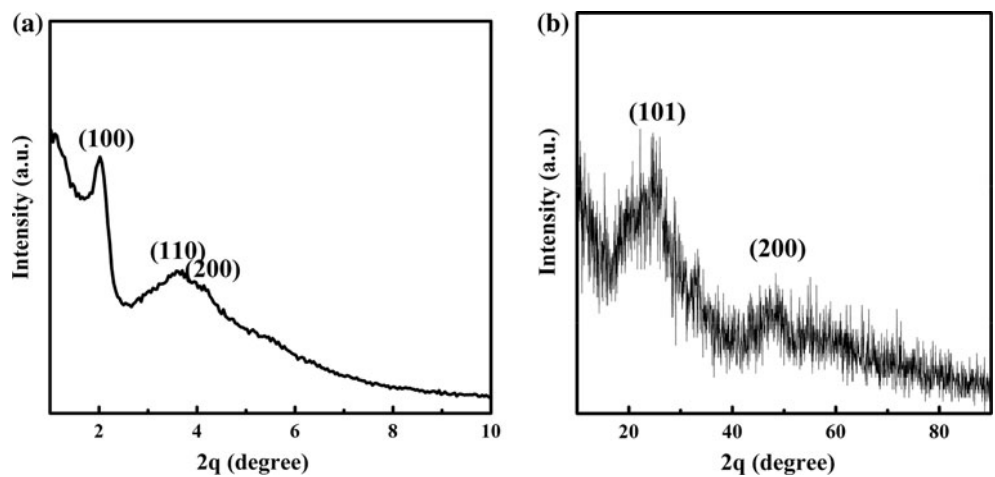

**Fig. 2** Low-angle (**a**) and wide-angle (**b**) XRD pattern of Hex-ncTiO$_2$/CTAB Nanoskeleton

Nanostructured TiO$_2$ samples with different crystallinity were obtained via acid-assisted hydrothermal treatment using the Hex-ncTiO$_2$/CTAB nanoskeleton as starting materials. Figure 3 presents the XRD patterns of the samples after hydrothermal treatment in HCl solutions of different concentration at 150°C for 24 h. The phase composition and purity of all the samples had been identified from the XRD patterns in Fig. 3. The peak locations and relative intensities for TiO$_2$ are cited from the Joint Committee on Powder Diffraction Standards (JCPDS) database. The peaks located at 25.4, 37.8, 48.0, 54.5° respond to the (101), (004), (200), (105 and 211) planes of the anatase phase (JCPDS 21-1272), and the peaks located at 27.5, 36.1, 54.4° respond to the (110), (101), (211) planes of the rutile phase (JCPDS 21-1276), respectively. All the products had been confirmed to be primarily anatase or a mixture of anatase and rutile.

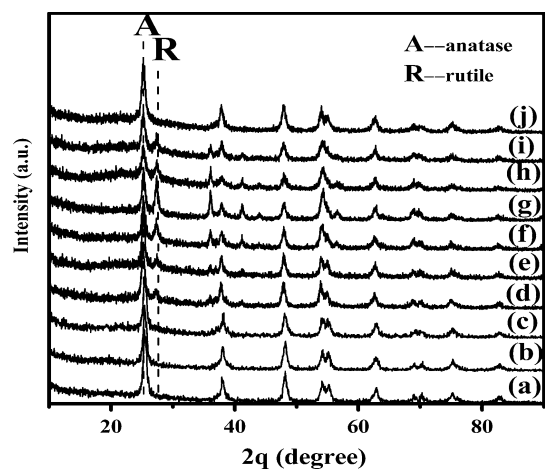

**Fig. 3** XRD patterns of hydrothermally synthesized TiO$_2$ products in different HCl solutions. **a** 0.1 M **b** 0.5 M **c** 1 M **d** 2 M **e** 3 M **f** 4 M **g** 5 M **h** 6 M **i** 7 M **j** 8 M

The anatase phase content for all the products had been calculated from the XRD patterns, using the following equation: $X_a = [1 + 1.26(I_r/I_a)]^{-1}$ [5], where $X_a$ is the share of anatase in the mixture, while $I_a$ and $I_r$ are the integrated intensities of the (101) reflection of anatase and the (110) reflection of rutile. Figure 4 presents the HCl concentration dependence of the ratio of anatase in the samples hydrothermally synthesized. We can observe directly the phase evolution process of obtained $TiO_2$ nanostructures with increased HCl concentrations. As the HCl concentration was increased from 0.1 to 1 M, the obtained products were found to be pure anatase phase in Fig. 3a–c. It presented a calculated result of 100% for the content of anatase in 0.1–1 M HCl solutions. With the increasing concentration of HCl ranged from 2 to 7 M, a mixture of anatase and rutile was obtained as showed in Fig. 3d–i. The content of anatase decreased and the content of rutile increased in the range of 1–5 M HCl solutions. A maximum of content of rutile can be obtained in the 5 M HCl solutions. Then the content of anatase increased and the content of rutile decreased in the range of 5–8 M HCl solutions. Finally 100% pure anatase phase sample was obtained again in 8 M HCl. XRD analysis indicated that the HCl concentration may play a key role during the crystalline phases formation of $TiO_2$ nanostructures.

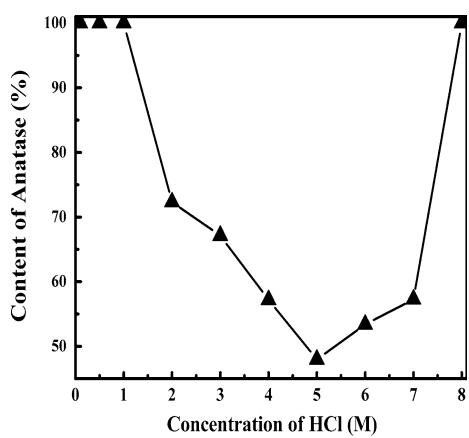

**Fig. 4** HCl concentration dependence of the ratio of anatase in the samples hydrothermally synthesized

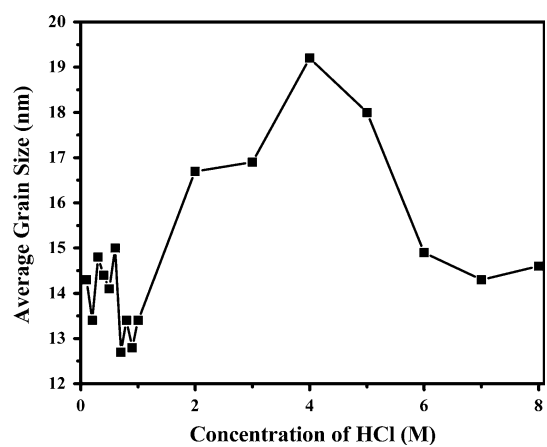

**Fig. 5** Average grain sizes of the products hydrothermally synthesized with different HCl concentrations

Average crystallite sizes of all the products were estimated using Scherrer equation: $D = 0.89\lambda/(\beta\cos\theta)$, where $\lambda$ is the employed X-ray wavelength, $\theta$ is the diffraction angle of the most intense diffraction peak, and $\beta$ is the full width at half maximum of the most intense diffraction peak (FWHM) [25]. Figure 5 shows the average grain sizes of the $TiO_2$ samples hydrothermally synthesized with increased HCl concentrations. The average crystallite size varied between 13 nm and 15 nm for the HCl concentration <1 M and increased slightly for the HCl concentration >1 M and achieved the maximum value of about 19 nm at 4 M HCl. Then the average crystallite size decreased for HCl concentration >4 M and finally reached the value of 14.5 nm for 8 M HCl. Moreover, it should be noted that the average crystallite size of pure anatase phase was smaller than mixture phase. Obviously, this related to crystalline phase, as increasing content of the rutile phase (Fig. 4), the average grain sizes increasing, by contraries, the average grain size decreasing with decreasing content of the rutile phase.

Figure 6 presents the TEM and SAED results of the $TiO_2$ nanostructures. The morphologies of $TiO_2$ samples changed dramatically with changing the concentrations of the HCl solutions. First, irregular size of aggregated nanoparticles with a mean particle diameter of 18 nm was observed in Fig. 6a for HCl concentration range below 1 M. The HRTEM image in Fig. 6g shows the lattice image with a lattice spacing of 0.352 nm that corresponds to the (101) lattice plane of anatase phase. The corresponding SAED pattern (inset image in Fig. 6a) indicated that the nanoparticles are polycrystalline anatase structure, in good agreement with the XRD results.

Figure 6b–e shows the $TiO_2$ nanostructures composed of aligned nanorods and irregular nanoparticles obtained with further increase in the HCl concentration from 2 to 7 M. The TEM results show that the $TiO_2$ nanoparticles are of irregular shape with an average size of 18 nm. The aligned nanorods maintained the analogous morphology with a width of around 50 nm and lengths of up to 300 nm in the concentration range from 2 to 7 M. Figure 6h presents the HRTEM investigations into the irregular nanoparticles and aligned nanorods. The lattice images of nanoparticles and nanorods were clearly observed, which indicated that these nanoparticles and nanorods had high degrees of crystallinity and phase purity. From the distance between the adjacent lattice fringes, we can assign the lattice plane on the nanoparticles and nanorods. The nanoparticles showed lattice spacing of $d = 0.354$ nm for the (101) plane of the anatase phase. The distance between the lattice fringes ($d = 0.325$ nm) in the aligned nanorods can be assigned to the interplanar distance of rutile phase (110) plane, which is well consistent with XRD results.

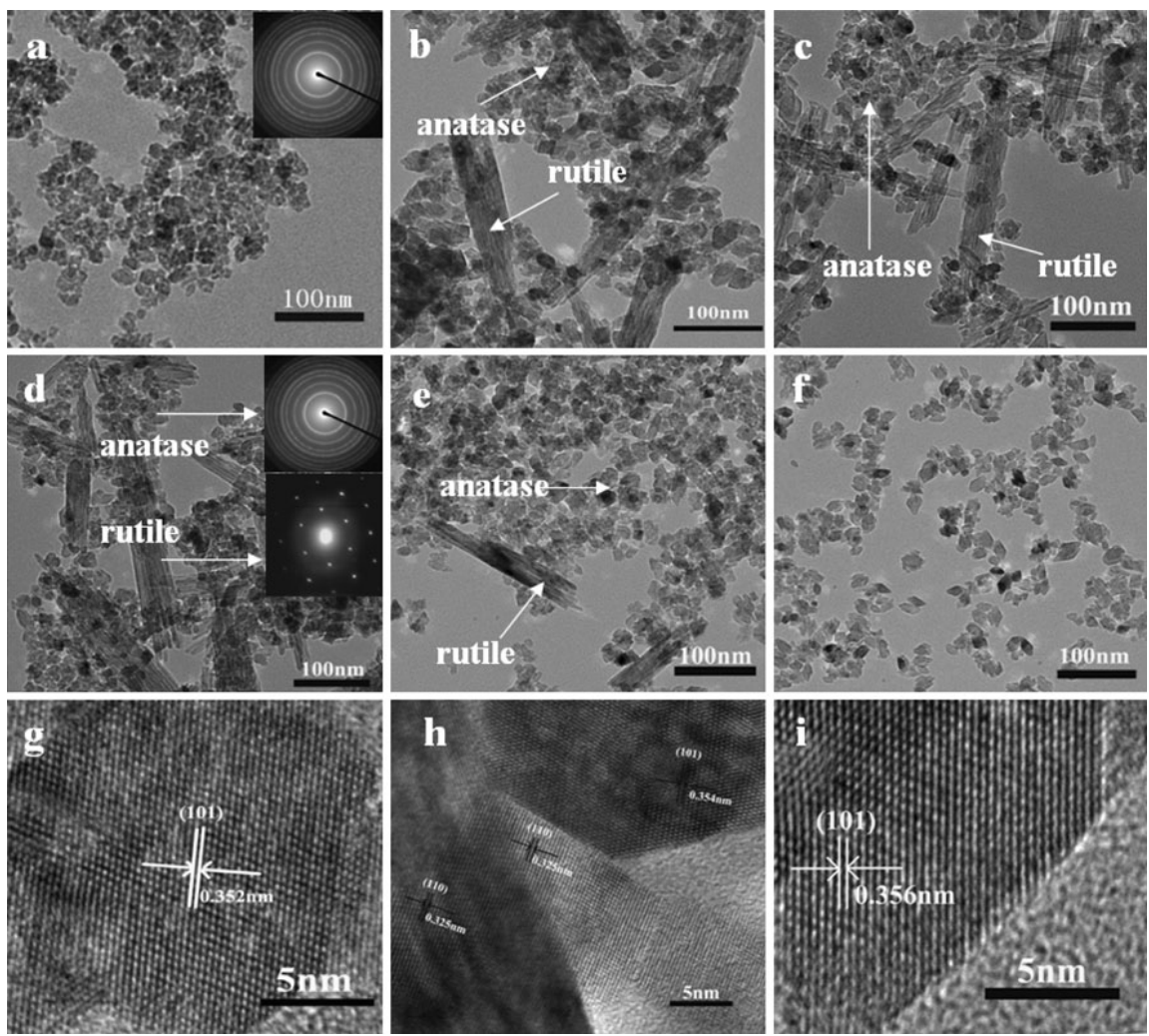

**Fig. 6** TEM images and corresponding SAED patterns (*inset*) of TiO$_2$ nanostructures hydrothermally synthesized in different HCl solutions: **a** 1 M **b** 3 M **c** 4 M **d** 5 M **e** 7 M **f** 8 M. Images of **g h i** are the HRTEM images for 1 M, 4 M, 8 M sample, respectively

Further observation by SAED (inset image in Fig. 6d) confirmed that the nanoparticles had a polycrystalline anatase structure, and the aligned nanorods were single crystalline TiO$_2$ with rutile structure.

Figure 6f shows that nearly monodispersed diamond-shaped nanocrystals with an average size of about 14 nm are formed as increasing the HCl concentration to 8 M. Further HRTEM analysis in Fig. 6i shows that the lattice fringes with an in interlayer distance of 0.356 nm is close to the 0.352 nm lattice spacing of the (101) planes in anatase TiO$_2$, which is in accordance with XRD results. From the TEM and SAED analysis, it can be concluded that different HCl concentrations affect not only the crystalline phase and crystallinity but also the morphologies of TiO$_2$ nanostructures. In addition, it can be noticed that the ratio of the rutile to anatase in the products increases with increasing HCl concentration range from 1 M to 5 M, and reach a maximum at 5 M, and decreases to zero with further increasing HCl concentration from 6 to 8 M, which corresponds to the XRD results in Fig. 3 and Fig. 4.

From the above results, the formation of TiO$_2$ nanostructures with various crystalline phases and morphology from the starting materials involved different nucleation and growth processes under the hydrothermal conditions. Two formation mechanisms have been proposed for the hydrothermal reaction [26–28]. One is the dissolution and recrystallization mechanism and the other is the in situ transformation mechanism. XRD analysis in Fig. 2 presented that the Hex-ncTiO$_2$/CTAB nanoskeletons as starting materials were mixture of poorly crystallized anatase and amorphous titania. It is expected that the reaction progresses through in situ transformation mechanism for the TiO$_2$ samples obtained in HCl solutions ranged from 0.1 to 7 M. The anatase nanocrystals in Hex-ncTiO$_2$/CTAB nanoskeletons may act as "seeds" for the growth of larger anatase nanoparticles. The transformation of amorphous

TiO$_2$ in the Hex-ncTiO$_2$/CTAB nanoskeleton exists a competition between the two growth units of rutile and anatase. Both anatase and rutile can grow from the [TiO$_6$] octahedra, and the phase formation proceeds by the structural rearrangement of the octahedral [13, 26–28]. During the process of TiO$_2$ crystal growth, HCl worked like a chemical catalyst to cause a change in the crystallization mechanism and decreased the activation energy for the rutile formation [29]. The growth of existed anatase nanocrystals and transformation of amorphous titania to anatase at low HCl concentration results in the formation of irregular nanoparticles with anatase phase in the 0.1 to 1 M HCl solutions. The increased HCl concentration ranged from 2 to 7 M, Cl$^{-1}$ can affect the O–Ti–O bonding structure and favor formation of the rutile nanorods structure from amorphous titania, which is consistent with some reports of rutile nanorods fabrications in acidic solution [24, 28]. Moreover, our hydrothermal experiments processed in higher concentration HCl solutions were gradually inclined to the dissolution and recrystallization mechanism. The rutile content decreased and the crystallite size of anatase decreased with the increasing HCl concentration ranging from 5 to 8 M as shown in Fig. 4 and Fig. 5. The solubility of titania oxides increased in the high acid solutions with the HCl concentration increasing steadily. The amorphous titania and anatase nanocrystals in the starting materials decomposed and recrystallized to form anatase nuclei according to the dissolution recrystallization mechanism. The observation of smaller uniform diamond-shaped nanoparticles with high quality single-crystal anatase structure in 8 M HCl solution clearly showed that dissolution–recrystallization process occurred with the high HCl concentration.

Anatase and rutile are two primary crystalline phases of TiO$_2$. The absorption onsets of anatase and rutile are located at about 387 nm and 413 nm, corresponding to band energy of 3.2 and 3.0 eV, respectively [1]. The valance band of anatase and rutile is mainly composed of O$2p$ states, while the conduction band is mainly formed of Ti$3d$ states. The band gap of TiO$_2$ is determined by the positions of conduction band and valance band, which is strongly related with its crystal structure, phase composition, grain size, and morphology. Therefore, the band gap of the mixture of anatase and rutile is between the values of pure anatase and rutile.

The UV–visible absorption spectra of all products hydrothermally synthesized in HCl solutions of different concentrations are shown in Fig. 7. The samples treated with 2–7 M HCl exhibited more or less red shift when compared with those treated with 1 and 8 M HCl. It is known that the relationship between the absorption band edge ($\lambda$) and the band gap ($E_g$) is shown as: $E_g$ (eV) = 1239.8/$\lambda$. The inset image in Fig. 7 shows the calculated band gaps for the different TiO$_2$ nanostructures using the equation. When the HCl concentration was lower than 1 M, the band gap of the TiO$_2$ nanoparticles varied over a narrow range from 2.99 to 3.01 eV resulting from pure anatase phase. The band gap of TiO$_2$ decreased to about 2.94 eV resulting from mixture anatase and rutile phase by increasing HCl concentration from 1 to 6 M. As increasing the HCl concentration to 8 M, the band gap increased to about 3.02 eV for 8 M, resulting from the formation of pure anatase phase as well. The band gap results were in well agreement with the crystalline phase identified by XRD. According to quantum size effect, the nanocrystals with larger size presented lower energy and displayed red shift. Conversely, the nanocrystals with smaller size displayed blue shift because of their higher energy. In our experiments, a red shift was observed for the products consisting of rutile nanorods compared with pure anatase particles, because the size of rutile nanorods was larger than anatase particles.

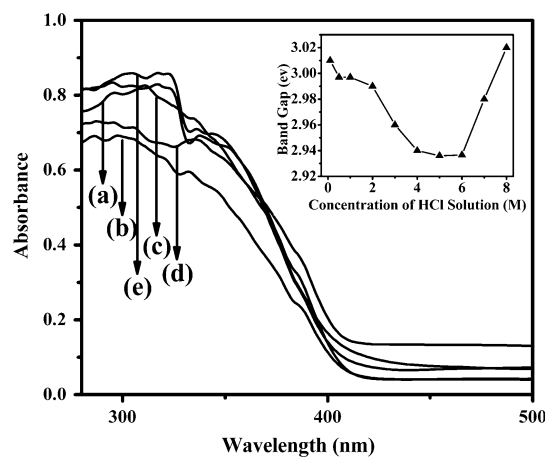

**Fig. 7** UV-vis spectra of the products hydrothermally synthesized in different HCl solutions: **a** 1 M **b** 3 M **c** 5 M **d** 7 M **e** 8 M. The inset shows the calculated band-gap energies

## Conclusions

Highly crystalline TiO$_2$ nanostructures were prepared through an acid-assisted hydrothermal process of the Hex-ncTiO$_2$/CTAB nanoskeleton. The HCl concentrations affected not only the crystalline phase but also the morphologies of TiO$_2$ nanostructures. Pure anatase nanoparticles were obtained in the lower HCl concentration range (0.1–1 M) and 8 M HCl, while a mixture of rutile nanorods and anatase nanoparticles were obtained for a broader concentration range of 2 M to 7 M. Different mechanisms were proposed for the phase formation and morphology changes of TiO$_2$ nanostructures with various HCl concentrations.

**Acknowledgements** This work was supported by the National Natural Science Foundation of China (Grant No. 20903034, 10874040) and the Cultivation Fund of the Key Scientific and Technical Innovation Project, Ministry of Education of China (Grant No. 708062).